\documentclass[runningheads]{llncs}

\usepackage[T1]{fontenc}
\usepackage{graphicx}
\usepackage{natbib}
\usepackage{amsmath,amssymb,amsfonts}
\usepackage{algorithmic}
\usepackage{graphicx}
\graphicspath{{figures/}}
\usepackage{textcomp}
\usepackage{balance}
\usepackage{multirow}
\usepackage{hyperref}
\usepackage{longtable}
\usepackage{enumitem}
\usepackage{xcolor}
\def\BibTeX{{\rm B\kern-.05em{\sc i\kern-.025em b}\kern-.08em
    T\kern-.1667em\lower.7ex\hbox{E}\kern-.125emX}}
\begin{document}
\title{Prompt Engineering Guidelines for Using Large Language Models in Requirements Engineering}

\author{Krishna Ronanki\inst{1,2}\orcidID{0009-0001-8242-6771} \and
Simon Arvidsson\textsuperscript{*}\inst{1,2} \and
Johan Axell\textsuperscript{*}\inst{1,2} \\
\thanks{These authors contributed equally to this work.}}
\authorrunning{Ronanki et al.}

\institute{Chalmers University of Technology, Gothenburg, Sweden \and
University of Gothenburg, Gothenburg, Sweden}
\maketitle              
\begin{abstract}
The rapid emergence of generative AI models like Large Language Models (LLMs) has demonstrated its utility across various activities, including within Requirements Engineering (RE). Ensuring the quality and accuracy of LLM-generated output is critical, with prompt engineering serving as a key technique to guide model responses. However, existing literature provides limited guidance on how prompt engineering can be leveraged, specifically for RE activities. The objective of this study is to explore the applicability of existing prompt engineering guidelines for the effective usage of LLMs within RE. To achieve this goal, we began by conducting a systematic review of primary literature to compile a non-exhaustive list of prompt engineering guidelines. Then, we conducted interviews with RE experts to present the extracted guidelines and gain insights on the advantages and limitations of their application within RE. Our literature review indicates a shortage of prompt engineering guidelines for domain-specific activities, specifically for RE. Our proposed mapping contributes to addressing this shortage. We conclude our study by identifying an important future line of research within this field.

\keywords{Requirements Engineering \and Generative AI \and Large Language Models \and Prompt Engineering \and Guidelines.}
\end{abstract}
\section{Introduction}

The development of Artificial Intelligence (AI) techniques such as Natural Language Processing (NLP), Machine Learning (ML) and Deep Learning (DL) methods for various requirements engineering (RE) activities, including requirements classification, prioritisation, tracing, ambiguity detection, and modelling has been increasing~\cite{alhoshan2023zero, zhao2021natural}. It has been empirically observed that natural language is the most commonly used medium for drafting requirements in the industrial context. This significant synergy between natural language and requirements has led to the emergence of NLP in the field of RE~\cite{zhao2021natural}. Despite the advancements in NLP for RE, there is a noticeable lack in the application of natural language generation (NLG)-based method in support of RE activities~\cite{zhao2021natural}. Moreover, the majority of existing state-of-the-art approaches for supporting RE activities and processes rely on ML/DL. These methods necessitate substantial amounts of task-specific labelled training data.

One potential solution to overcome the need of large quantities of high quality labelled training data and advance NLG within RE is using pre-trained generative AI models like Large Language Models (LLM). Utilising LLMs for performing RE activities can potentially remove the need for large amounts of labelled data. Recent LLMs are demonstrating increasingly impressive capabilities when performing a wide range of tasks~\cite{brown2020language}, including their significant enhancements in NLP tasks~\cite{haque2022think}. The GPT-3.5 LLM has demonstrated a surprising proficiency in specific technical tasks~\cite{choi2023chatgpt}, including RE~\cite{ReqClassificationZS}.

However, the input prompts to the LLM are observed to significantly impact the quality of the LLM's output~\cite{10629163}. The emerging practice of utilising carefully selected and composed natural language instructions to achieve a desirable output from an LLM is called \emph{prompt engineering}~\cite{10.1145/3491102.3501825}. Prompt engineering plays a crucial role as the selection of prompts can have a significant impact on downstream tasks~\cite{perez2021true}. However, there are no guidelines or strategies that RE practitioners can leverage to utilise LLM effectively.

To address this research gap, we conducted a systematic review of primary studies to identify and analyse existing prompt engineering guidelines for LLMs and explore their applicability within RE. By mapping these guidelines to RE activities, we aim to bridge a gap in the literature and provide a structured approach for both researchers and practitioners. In order for us to realise these objectives, the following research questions were formed:

\noindent\fbox{
\begin{minipage} {.96\columnwidth}
\textbf{RQ1:} What are the existing prompt engineering guidelines for effective use of LLMs?

\textbf{RQ2:} What are the advantages and the limitations of using these guidelines for the effective usage of LLMs within RE?

\textbf{RQ3:} What are the relevant prompt engineering guidelines for the effective usage of LLMs within RE?
\end{minipage}
}\\

Section~\ref{sec:bg} present the relevant background literature that motivates our research goals. Section~\ref{sec:methods} describes our research design and execution. Section~\ref{sec:results} presents the results of the our study and answer the \textbf{RQ1}, \textbf{RQ2} and \textbf{RQ3}. We discuss the implications of the gathered results along with the threats to validity of our study in Section~\ref{sec:discussion}. Finally, in Section~\ref{sec:conclusion}, we conclude our study and present potential areas for further research based on the discussed implications. 

\section{Background and Related Work} \label{sec:bg}

Recent research in prompt engineering has proposed guidelines for effective prompt engineering of generative AI models, particularly LLMs and large vision models~\cite{DesignGuidelinesGenerativeModels}. The existing research on the development of prompting techniques for domain-specific tasks such as RE, however, remains sparse with some generally applicable outcomes provided by White et al.~\cite{PromptPatterns}. The literature on prompt engineering guidelines and approaches tied specifically to RE remains limited at the time of this study, with only a few papers partially covering guidelines for tasks and prompt learning within RE activities.

Rodriguez et al.~\cite{10260721} explore the process of prompt engineering to extract link predictions from an LLM. They provide detailed insights into their approach for constructing effective prompts and propose multiple strategies for leveraging LLMs. Their key takeaways include the impact of minor changes to prompts on the quality of the outputs, the effect of chain-of-thought reasoning, and the importance of specifying targeted usage to get desired outcomes. However, this study's biggest drawback is its limitation to only one task.

Ronanki et al.~\cite{Ronanki2024} evaluate the effectiveness of 5 of the prompt patterns’~\cite{white2023chatgpt} ability to make GPT-3.5 turbo perform binary requirements classification and requirements traceability tasks. They also offer recommendations on which prompt pattern to use for a specific RE task and provide an evaluation framework as a reference for researchers and practitioners who want to evaluate different prompt patterns for different RE tasks. The results of this study are limited to only the two mentioned RE tasks and the five prompt patterns that were used in the experiments.

Sasaki et al.~\cite{10633588} present a systematic literature review focusing on practical applications of prompt engineering in SE. They identified and classified prompt engineering patterns, showcasing their application across different SE tasks. They also provide a taxonomy of prompt engineering techniques for SE that has five categories, eleven sub-categories and twenty one patterns along with the problems they can applied to and the solution you receive. However, the prompt engineering patterns covered within the study are from various SE tasks like code generation, testing, debugging, etc. 

The reviewed studies broadly agree that the ``context'' and ``structure'' of prompts significantly impact the quality of generated outputs. However, there is considerable variation in the recommended approaches for achieving high-quality results. ``Few-shot prompting'' has been observed to improve the LLM output quality compared to alternative methods~\cite{CoTImproveLLMsZs, FSPrompts}. Conversely, other studies highlight cases where few-shot prompting falls short, suggesting that its effectiveness depends on specific conditions~\cite{AnalyticalTemplate, ThinkStepByStepExtended}.

Furthermore, research on models like GPT-3, GPT-3.5, and ChatGPT indicates persistent challenges in tasks requiring emotional understanding or mathematical reasoning, regardless of the prompting approach used~\cite{EmotionEvaluations}. Additionally, these models are prone to confidently generating incorrect facts, a phenomenon known as hallucinations. Various studies propose mitigation strategies to address these inaccuracies~\cite{ReasonStepByStep}. Recent studies have also explored applying prompt engineering guidelines to new domains, including SE. These studies often propose domain-specific adaptations of existing guidelines, expanding the applicability of prompt engineering techniques~\cite{PromptPatterns, AnalyticalTemplate, ThinkStepByStepExtended, CoTImproveLLMsZs, EmotionEvaluations}. These sources, while crucial for shaping and enhancing our understanding of prompt engineering for LLMs, fail to address the gap of lack of standard prompt engineering guidelines applicable for RE activities, further strengthening the motivation of our study.

Through the works of Pressman, R.S.~\cite{pressman2005software} and Sommerville, I.~\cite{10.5555/549198}, the five main RE activities identified in the literature are elicitation, analysis, specification, validation, and management. These activities are crucial for successful product development and face various challenges. Our mapping of prompt engineering guidelines to RE was done while focusing primarily on these five activities.

\section{Methodology} \label{sec:methods}

We conducted a systematic review of primary studies to identify prompt engineering guidelines for LLMs to answer \textbf{RQ1}. Then, we conducted interviews with three RE experts to gain insights into how the extracted guidelines can be used to leverage LLMs more effectively during RE activities. These interviews helped us gain insight into the advantages and limitations of the prompt engineering guidelines, answering \textbf{RQ2}. Based on the expert's insights, a mapping of relevant prompt engineering guidelines to different RE activities was performed following the thematic synthesis approach~\cite{CRUZES2011440}. This mapping, has the potential to help RE practitioners leverage LLMs more effectively, and answers \textbf{RQ3}.

\subsection{Systematic Review of Primary Studies}

We based our systematic review process partially on the guidelines for conducting a systematic literature review~\cite{SLRGuidelines}. To construct a comprehensive search strategy, we identified relevant keywords through prior research in related domains and a snowballing approach. Based on our research objectives and questions, we formulated the following search string: (``Prompt Engineering'' OR ``Prompt Patterns'' OR ``Prompt Design'' OR ``Prompt Catalog'' OR ``Prompt Guidelines'') AND (``Large Language Models'' OR ``Generative AI''). During the initial exploration, we observed inconsistencies in terminology across studies. Consequently, we expanded the search string to improve coverage. The string was then adapted to match the interface requirements of each database while preserving the original keywords. 

We restricted our search to papers published from 2018 onward, as this marks the introduction of transformer architectures with OpenAI’s Generative Pre-trained Transformer (GPT) and Google’s Bidirectional Encoder Representations from Transformers (BERT)~\cite{TransformerArchitecture, BERTpresentation, FirstGPTPresented}. This time frame constraint also helps ensure relevancy and state-of-the-art within the field of prompt engineering to match the rapid development of large generative models.

Our review covered five databases: ACM, Scopus, IEEE Xplore, ScienceDirect, and arXiv. The first four were selected due to their extensive coverage and credibility in computer science and software engineering research. arXiv was later included to capture emerging studies, given the field's fast-paced development.

To identify relevant primary studies for this systematic review, we applied inclusion and exclusion criteria, as detailed in Table~\ref{tab:criteria}. Each study was initially assessed based on its title, abstract, introduction, conclusion, and keywords. If ambiguity remained after this stage, a full-text review was conducted to determine its relevance. Applying the search string to the selected databases yielded 271 studies for initial screening. After evaluating them against the inclusion and exclusion criteria presented in Table~\ref{tab:criteria} and removing duplicates, 28 studies (10.3\%) were identified as primary studies.

\begin{table}[ht!]
\centering
\resizebox{\columnwidth}{!}{
\begin{tabular}{|l|l|}
\hline
\textbf{Inclusion criteria} & \textbf{Exclusion criteria}
\\ \hline
Written in English language & \begin{tabular}[c]{@{}l@{}}Papers with sections or content\\ in languages other than English\end{tabular} \\ \hline
Date of publication from 2018 & Published prior to 2018                                  \\ \hline
Emphasises generative AI models & Unrelated to generative AI models                           \\ \hline
Focus on Natural language prompts & Emphasis on model tuning                                \\ \hline
Relevant to RQ1 & Does not contain prompt engineering guidelines                                     \\ \hline
\end{tabular}
    }
\caption{Inclusion and exclusion criteria for the review.}
\label{tab:criteria}
\end{table}

In order to conduct extraction of relevant data, the form depicted in Table~\ref{tab:DataExctraction} was used. This form was established in order to ensure an organised as well as standardised data collection process. It also facilitated the process of distinguishing relevant data for our research question. 

\begin{table}[ht!]
\resizebox{\columnwidth}{!}{
\begin{tabular}{|l|p{7.5cm}|l|}
\hline
\textbf{Data} & \textbf{Description}                                                          & \textbf{Relevant RQ} \\ \hline
Year          & To ensure temporal relevance of   the extracted prompt engineering guidelines & General              \\ \hline
Model Type    & Text-to-image and text-to-text.                      & RQ1                  \\ \hline
Prompt Method & What prompt engineering   techniques were studied?                            & RQ1                  \\ \hline
Guidelines    & What guidelines are presented?                                                & RQ1                  \\ \hline
Findings      & Strengths or limitations of the   guidelines presented                        & RQ1                  \\ \hline
\end{tabular}
}
\caption{Data extraction form}
\label{tab:DataExctraction}
\end{table}

\subsection{Interviews: Data Collection and Analysis}

To assess the advantages and limitations of the identified prompt engineering guidelines and gain insights on which guidelines are relevant for the activities involved within the selected RE phases, we conducted semi-structured interviews~\cite{1130282271686871424} with three experts from different academic institutions, each specialising in distinct areas of RE. The primary selection criterion was expertise in RE, with additional consideration given to knowledge in AI4RE and related fields. Experts were identified based on their academic background and publication history. While our study focuses on RE, the selected experts provided diverse perspectives, spanning fields such as RE4AI and NLP4RE as well. Experts were contacted via email with an introduction to the study and details on the interview process. All interviews were conducted online via Zoom, each lasting approximately 30 minutes. We ensured all the interviews were conducted by following the ethical interview checklist presented by P.E. Strandberg~\cite{8870192}.

We chose qualitative content analysis~\cite{drisko2016content} to extract insights from the qualitative data collected from the interviews. The transcripts of the interviews were edited, validated through member checking~\citep{799955}, and anonymised before we performed content analysis. We then conducted a thematic synthesis to map the relevant prompt engineering guidelines to different activities within the selected RE phases. Our thematic synthesis approach was inspired by Cruzes and Dybå~\cite{CRUZES2011440}, involved summarising, integrating, and comparing findings on a specific topic to generate new insights, such as theories, frameworks, or conclusions. The final mapping was developed by integrating the insights provided by the RE experts for mapping of the guidelines to the RE activities.

\section{Results} \label{sec:results}

The results of this study are presented in three subsections, each corresponding to an \textbf{RQ}. Subsection~\ref{subsec:RQ1} addresses \textbf{RQ1}, presenting a non-exhaustive list of prompt engineering guidelines. These guidelines are categorised into the nine most frequently occurring themes identified in our review. Subsection~\ref{subsec:RQ2} explores the advantages and limitations of applying the identified prompt engineering guidelines in RE based on expert interviews, providing insights necessary for answering \textbf{RQ2}. Subsection~\ref{subsec:RQ3} answers \textbf{RQ3} by mapping the identified guidelines to relevant RE activities.

\subsection{Prompt Engineering Guidelines} \label{subsec:RQ1}

From the 28 included primary studies, we identified and extracted 36 prompt engineering guidelines. These guidelines were categorised into nine distinct themes based on their defining characteristics. Each theme was defined to capture the underlying concepts and group similar guidelines, providing a clearer overview of their overall intent and purpose.

\subsubsection{Context}
The ``Context'' theme revolves around guidelines that relate to contextual information in any manner for prompts.
\subsubsection{Persona}
The ``Persona'' theme is a high-level abstraction of guidelines that revolves around strategies for LLMs to take on specific or different perspectives on specified tasks by using prompts. Persona in this instance can be compared to perspectives or points of view.
\subsubsection{Templates}
The ``Templates'' theme encapsulates guidelines that only provide an explicit structure of prompts, known as a template.
\subsubsection{Disambiguation}
The ``Disambiguation'' theme refers to guidelines that aim to address ambiguity, clarification, or understanding of intent.
\subsubsection{Reasoning}
The ``Reasoning'' theme captures guidelines that aim to affect reasoning capabilities or the ability to think through complex problems or tasks in a generated output.
\subsubsection{Analysis}
The theme ``Analysis'' revolve around guidelines examining, evaluating, or analysing information or tasks. 
\subsubsection{Keywords}
The theme ``Keywords'' represent guidelines that involve any use of single-word modifiers to prompts.
\subsubsection{Wording}
The theme ``Wording'' refers to guidelines that relate to choices of words, text formatting, writing styles, or inclusion and exclusion of text.
\subsubsection{Few-shot Prompts}
The theme ``Few-shot Prompts'' categorises guidelines that are intended for any form of few-shot prompting.

\begin{figure}[t]
    \centering \includegraphics[width=9cm]{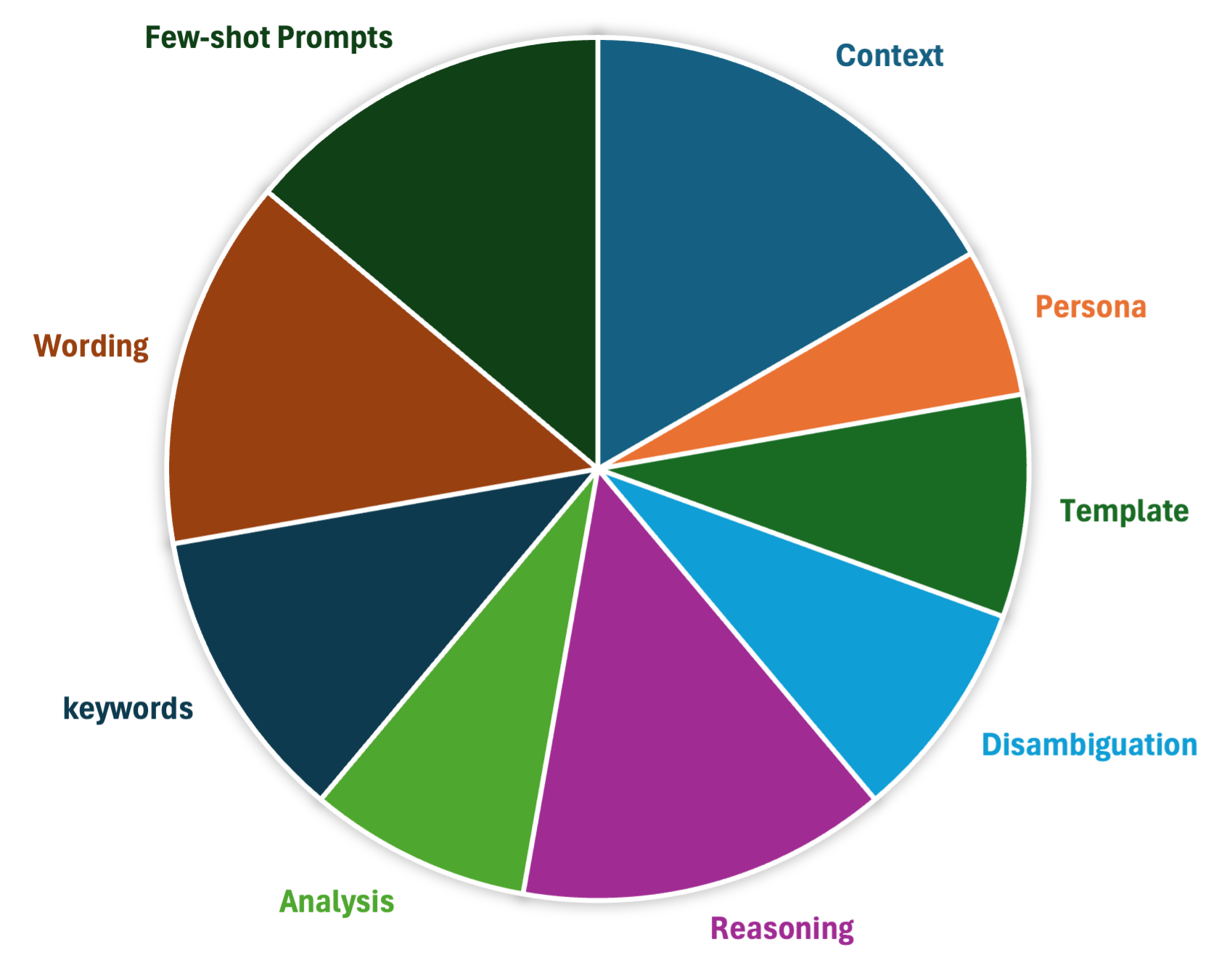}
    \caption{Theme distribution of the extracted guidelines from the studies in the review.}
    \label{fig:ShareOfThemes}
\end{figure}

Each guideline was assigned to one of the nine themes and given a unique identifier, as listed in Table~\ref{tab:PEThemes}. This categorisation provides insight into the most common types of guidelines found in the literature. As shown in Fig.~\ref{fig:ShareOfThemes}, the ``Context'' theme consists of the highest number of guidelines, followed by ``Reasoning'', ``Wording'', and``Few-shot Prompts'' themes. The ``Persona'' theme had the lowest number of guidelines, with only 2. To reference specific guidelines in this paper, we use a combination of the theme's initial letter and the guideline's identifier. For example, C1 refers to the first guideline within the ``Context'' theme.

\begin{table*}[h!]
    \centering
    \resizebox{\textwidth}{!}{
    \begin{tabular}{|c|c|}
        \hline
        \textbf{Theme} & \textbf{Description} \\ 
        \hline
        \multicolumn{1}{|c|}{\begin{tabular}[c]{@{}c@{}}\textbf{Context}\\ \textbf{(C)}\end{tabular}}  & \multicolumn{1}{l|}{\begin{tabular}[c]{@{}p{14.25cm}@{}}1. Adding context to examples in prompts produce more efficient and informative output. \cite{EfficientAndInformative} \\ 2. Provide context to all prompts to avoid output hallucinations. \cite{ContextHallucinations}\\ 3. Provide context of the prompt to ensure a closely related output. \cite{ContextRelatedOutput,ContextTokens} \\ 4. The more context tokens pre-appended to prompts, the more fine-grained output.\cite{ContextTokens} \\ 5. The Context Manager Pattern enables users to specify or remove context for a conversation with an LLM \cite{white2023prompt} \\ 6. Providing more context and instructions is an effective strategy to increase the semantic quality of the output~\cite{10629163}\end{tabular}}
        \\ 
        \hline
        \multicolumn{1}{|c|}{\begin{tabular}[c]{@{}c@{}}\textbf{Persona}\\ \textbf{(P)}\end{tabular}} & \multicolumn{1}{l|}{\begin{tabular}[c]{@{}p{14.25cm}@{}} 1. Improves the generation quality by conditioning the prompt with an identity, such as ``Python programmer'' or ``Math tutor'' \cite{GivePromptIdentity}\\ 2. To explore the requirements of a software-reliant system, include: \\ \hspace{\parindent} - ``I want you to act as the system'',\\ \hspace{\parindent} - ``Use the requirements to guide your behaviour''\cite{white2023chatgpt} \end{tabular}}
        \\  
        \hline
        \multicolumn{1}{|c|}{\begin{tabular}[c]{@{}c@{}}\textbf{Templates} \\ \textbf{(T)}\end{tabular}} & \multicolumn{1}{l|}{\begin{tabular}[c]{@{}p{14.25cm}@{}} 1. To improve reasoning and common sense in output, follow a template such as: \\ \hspace{\parindent} - ``Reason step-by-step for the following problem. [Original prompt inserted here]'' \cite{ReasonStepByStep} \\ 2. The following prompt template has shown an impressive quality of AI art: \\ \hspace{\parindent} - ``[Medium] [Subject] [Artist(s)] [Details] [Image repository support]''\cite{ArtPrompting} \\ 3. I am going to provide a template for your output; This is the template: PATTERN with PLACEHOLDERS \cite{white2023prompt}\end{tabular}}
        \\
        \hline
        \multicolumn{1}{|c|}{\begin{tabular}[p{1.6cm}]{@{}p{1.6cm}@{}}\textbf{Disambi-guation}  \\ \textbf{     (D)}\end{tabular}}& \multicolumn{1}{l|}{\begin{tabular}[c]{@{}p{14.25cm}@{}} 1. Ensure any areas of potential miscommunication or ambiguity are caught, by providing a detailed scope: \\ \hspace{\parindent} - ``Within this scope'',\\ \hspace{\parindent} - ``Consider these requirements or specifications'' \cite{white2023chatgpt}\\ 2. To find points of weakness in a requirements specification, consider including: \\ \hspace{\parindent} - ``Point out any areas of ambiguity or potentially unintended outcomes'' \cite{white2023chatgpt}\\ 3. The persona prompt method can be used to consider potential ambiguities from different perspectives. \cite{PromptPatterns}\end{tabular}}
        \\
        \hline
        \multicolumn{1}{|c|}{\begin{tabular}[c]{@{}c@{}}\textbf{Reasoning} \\ \textbf{(R)}\end{tabular}}& \multicolumn{1}{l|}{\begin{tabular}[c]{@{}l@{}} 1.  Prepending ``Let’s think step by step'' improves zero-shot performance. \cite{ThinkStepByStep}\\ 2. Extending the previously known ``Let's think step by step'', with ``to reach the right conclusion,'' to highlight decision-making in \\ the prompt. \cite{ThinkStepByStepExtended} \\ 3. Chain Of Thought (CoT) prompting improves LLM performance and factual inconsistency evaluation \\ compared to Zero-shot.\cite{CoTImproveLLMsZs,FactualInconsistency} \\ 4. Tree of thought (ToT) allows LLMs to perform deliberate decision making by considering multiple different reasoning paths \\ and self-evaluating choices to decide the next course of action \cite{yao2024tree} \\ 5. The intent of the Cognitive Verifier pattern is to force the LLM to always subdivide questions into additional questions that can \\ be used to provide a better answer to the original question \cite{white2023prompt}\end{tabular}}
        \\
        \hline
        \multicolumn{1}{|c|}{\begin{tabular}[c]{@{}c@{}}\textbf{Analysis} \\ \textbf{(A)}\end{tabular}} & \multicolumn{1}{l|}{\begin{tabular}[c]{@{}p{14.25cm}@{}} 1. self-consistency boosts the performance of chain-of-thought prompting \cite{wang2022self} \\ 2. The best of three strategy improves the LLM output stability and detecting hallucinations \cite{10.1007/978-3-031-48550-3_17}\\ 3. Emotion-enhanced CoT prompting is an effective method to leverage emotional cues to enhance the ability of ChatGPT on \\ mental health analysis. \cite{EmotionEvaluations}\end{tabular}}
        \\
        \hline
        \multicolumn{1}{|c|}{\begin{tabular}[c]{@{}c@{}}\textbf{Keywords}\\ \textbf{(K)} \end{tabular}} & \multicolumn{1}{l|}{\begin{tabular}[c]{@{}p{14.25cm}@{}} 1. When picking the prompt, focus on the subject and style keywords instead of connecting words. \cite{DesignGuidelinesGenerativeModels}\\ 2. Pre-appending keywords to prompts are shown to greatly improve performance by providing the language model with \\ appropriate context. \cite{PrependingContext}\\ 3. Modifiers/Keywords can be added to the details or image repository sections of a template such as:\\  \hspace{\parindent} - ``[Medium] [Subject] [Artist(s)] [Details] [Image repository support]'' \cite{ArtPrompting} \\ 4. The inclusion of multiple descriptive keywords tends to align results closer to expectations. \cite{OptimizingPromptsText2Img} \end{tabular}}
        \\
        \hline
        \multicolumn{1}{|c|}{\begin{tabular}[c]{@{}c@{}}\textbf{Wording} \\ \textbf{(W)}  \end{tabular}}& \multicolumn{1}{l|}{\begin{tabular}[c]{@{}p{14.25cm}@{}} 1. In translation tasks, adding a newline before the phrase in a new language increases the odds that the output sentence \\ is still English. \cite{EfficientAndInformative} \\ 2. A complete sentence definition with stop words performs better as a prompt than a set of core terms that were extracted \\from the complete sentence definition after removing the stop words. \cite{StopWords}\\ 3. Words such as ``well-known'' and ``often used to explain'' are successful for analogy generation. \cite{PromptAnalogy} \\ 4. Modifying prompts to resemble pseudocode tend to be the most successful in coding tasks. ~\cite{PseudocodePrompts, white2023chatgpt}\\ 5. Prompts to contain explicit algorithmic hints in engineering tasks perform better. ~\cite{PseudocodePrompts}\end{tabular}}
        \\
        \hline
        \multicolumn{1}{|c|}{\begin{tabular}[p{1.6cm}]{@{}p{1.6cm}@{}}\textbf{Few-shot Prompts}  \\ \textbf{       (F)}  \end{tabular}}& \multicolumn{1}{l|}{\begin{tabular}[c]{@{}p{14.25cm}@{}}1. Inclusion of ``Question:'' and ``Answer:'' improves the response, but rarely gives a binary answer. ~\cite{FSLegalPromptingQnA}\\ 2. For easier understanding, number examples in few-shot prompting. ~\cite{FSPrompts}\\ 3. The format of [INPUT] and [OUTPUT] should linguistically imply the relationship between them. ~\cite{FSPrompts}\\ 4. Specifications can be added to each [INPUT] and [OUTPUT] pair to give extra insight into complicated problems. ~\cite{FSPrompts} \\ 5.  In Few-shot prompting include a rationale in each shot (Input-rationale-output). ~\cite{RationaleFSprompts} \end{tabular}}
        \\
        \hline
    \end{tabular}}
    \caption{Thematic Classification of Prompt Engineering Guidelines}
    \label{tab:PEThemes}
\end{table*}

\subsection{Expert Insights: Interview Analysis}  \label{subsec:RQ2}

Interviews were conducted with three different RE experts, enquiring about their views on the advantages and the limitations of guidelines from our review and their possible usage for LLMs within RE activities.

\subsubsection{Requirements Elicitation}

Each of the experts interviewed emphasised the importance of context in prompts, especially for for requirement elicitation. Regarding the advantages, Expert 2 discussed the potential of leveraging contextual prompts for LLMs in this process, noting that they could encourage creative requirements exploration. They stated, ``It may be creative requirements, like [for the LLM] to say, hey, have you thought of this? And then a person says, oh, that's a good idea, or no, that's a bad idea.'' Expert 3 emphasised that templates could be beneficial across various prompt-related activities and RE activities where LLMs provide value, specifically mentioning requirements elicitation as a key application. They suggested that keyword-based guidelines can be appropriate for requirements elicitation.

However, the experts also had concerns regarding the effectiveness of term ``context'' included in the guidelines, arguing how it could be ambiguous as it is a quite general word. Expert 3 stated ``I would say that context may need to be decomposed in somehow because otherwise, the guideline may be too generic.'' Expert 1 mentioned that the context that the guidelines refer to may vary depending on assumptions made by different stakeholders. Expert 2 brought up another limitation, on how the context may not fully cover what the stakeholders want, stating ``So you can use it to elicit requirements, but they're not necessarily requirements that anyone wants to implement.''

\subsubsection{Requirements Analysis}

Expert 2 acknowledged a potential advantage in using LLMs to assess key qualities of requirements, stating, ``If you’re asking [the LLM], ‘Is this maintainable? Is this verifiable? Is this unambiguous?’ then you could come back and say yes or no, and that might be useful.'' They 2 also suggested that templates could be useful for analysing completeness of requirements.

However, they ultimately emphasised the limitations, explaining that the uncertainty surrounding the output, coupled with a lack of confidence in the feedback provided by LLMs, would make the results unreliable and of limited practical value. Expert 1 highlighted a key limitation, noting a lack of confidence in the feedback generated and the inherent uncertainty surrounding it. They raised concerns about the inconsistencies in LLM-generated outputs, particularly the variability in responses even when identical prompts are used. They cautioned against over-reliance on templates in requirements analysis.

Expert 1 also pointed out the difficulty in understanding how an LLM interprets reasoning steps. Similarly, Expert 2 stated, ``It is only capturing one very narrow type of reasoning. If you are trying to get it to reason in a process view, like step one, step two, step three, I am not even sure that is reasoning.'' They further questioned whether LLMs could effectively support the type of reasoning required in RE.

Despite these concerns, Expert 3 noted potential advantages, stating, ``It could be useful for the analysis of requirements towards the generation of system architecture'' They also suggested that embedding prompting guidelines within LLMs could support automated sub-tasks that aid in requirements analysis, referencing implementation approaches such as AutoGPT~\cite{agpt}. Overall, experts agreed that requirements analysis is likely too complex for current LLMs to handle effectively. They expressed scepticism about the capability of existing LLMs to fully support this task given their current state of development.

\subsubsection{Requirements Specification}

Expert 1 emphasised the potential to iteratively explore unclear aspects of requirements, stating, ``You can explore, maybe even iteratively, any unclear aspects of your requirements and, as I say, any ambiguities where there could be potential misunderstanding'' They further suggested that integrating persona-based guidelines could enhance this process by enabling ambiguity exploration from multiple perspectives. They stressed the importance of justifications, explaining that there is a need for reasoning behind why a requirement is deemed ambiguous or weak.

Expert 2 identified a key advantage, noting that ``it [LLM] could help to point out weaknesses in a requirement specification'' and that ``prompt engineering could help point out ambiguity.'' However, they also acknowledged that ambiguity might not always be a concern, referencing prior studies on smaller teams where such issues had minimal impact.

Expert 3 exclusively highlighted advantages, emphasising that these guidelines could be beneficial not only for requirement specification but also for requirements review. They provided an example related to guideline D2, stating, ``So this is a useful prompt for sure, it’s going to change the way we make requirements reviews.'' They further elaborated on the critical role of requirements reviews, particularly in safety-critical contexts.

\subsubsection{Requirements Validation}

Experts 1 and 3 acknowledged potential advantages of using LLMs for requirements validation. Expert 3 suggested that the use of LLMs in validation could be a promising area for further research. They proposed that LLMs could assist in validating requirements against the needs of specific user types, rather than solely acting as the system itself, stating,``Validating requirements based on the needs of a specific user could be beneficial, instead of simply acting as the system.'' However, they emphasised the necessity of providing extensive descriptions of the system to the LLM within the prompts as a potential limitation.

Expert 1 also noted a potential advantage in using LLMs to explore the validity of specific goals or targets from multiple perspectives. However, they raised concerns regarding the role of context among different personas. In this context, context refers to what the personas know about the system, and a key limitation is determining the appropriate scope of this knowledge. Expert 1 cautioned that restricting the LLM’s perspective to a specific dimension of the problem could significantly narrow the scope of validation, making it applicable only to the predefined perspective. They stated, ``If you limit the LLM only to look from a certain dimension or certain perspective on the problem, then you scope down your validation a lot to be only valid for that certain perspective that you define beforehand.'' They also acknowledged a key advantage of templates, explaining that templates are always nice because you can validate that templates actually work nicely beforehand, and then people can just use it without having to analyse if the template is good.

Expert 2 expressed scepticism regarding the use of LLMs for requirements validation, stating, ``I think it's a bad idea to use the prompt engineering for validation.'' They argued that LLMs lack the capability to determine whether a system is correct, making them unsuitable for requirements validation. Additionally, Expert 3 highlighted a fundamental challenge in using LLMs for validation, noting that validation is the final stage of the requirements process. They expressed uncertainty about whether current LLMs possess the necessary accuracy to process the substantial amount of technical information that systems typically require at this stage of development.

\subsubsection{Requirements Management}

Experts 2 and 3 identified potential advantages of keyword-based guidelines, particularly in their ability to assist with sorting and classifying requirements into categories. Expert 3 further suggested that these guidelines could support additional requirements management tasks, such as similarity analysis and traceability.

Expert 1 focused on the role of keywords in providing context, explaining that they could help establish assumptions and enhance the precision of LLM-generated outputs. However, they also highlighted a key limitation, noting that increased precision could ``limit the scope out of which the output can come'' potentially excluding important aspects from consideration. They also pointed to a broader advantage of structured reasoning within LLMs, emphasising that ``you get much better traceability or track why a certain decision has been made by the LLM or why a certain output was generated.'' This suggests that structured guidelines could enhance transparency in LLM-driven requirements analysis. Expert 3 questioned whether keywords alone would be sufficient for requirements management tasks, as these tasks often require in-depth project and domain knowledge.

\subsection{RE Activities and Applicable Guidelines}  \label{subsec:RQ3}

The proposed mapping between RE activities and the identified guideline themes, along with the corresponding justifications, is outlined in this section. A detailed representation of this mapping is provided in Table~\ref{tab:REGuidelinesMapping}. 

\begin{table}[ht!]
    \centering
    \resizebox{\columnwidth}{!}{
    \begin{tabular}{|l|p{8.5cm}|}
        \hline
        \textbf{RE Phase} & \textbf{Applicable Guidelines} \\ \hline
         Elicitation & Context, Template, Keyword\\
         \hline
         Analysis & Template, Analysis, Reasoning\\
         \hline
         Specification &  Persona, Disambiguation\\
         \hline
         Validation & Persona, Template\\
         \hline
         Management & Keyword, Reasoning\\
         \hline
    \end{tabular}
    }
    \caption{Mapping of Guidelines and RE}
    \label{tab:REGuidelinesMapping}
\end{table}

We believe the users performing requirements elicitation using LLMs can benefit from employing guidelines categorised under the ``context'', ``template'' and ``keyword'' themes, presented in Table~\ref{tab:PEThemes}. This mapping was done based on the experts' insights presented in detail within subsection~\ref{subsec:RQ2}. 

For requirements analysis, the ``template'', ``analysis'' and ``reasoning'' guidelines can be useful. Guidelines such as ``self-consistency''~\cite{wang2022self} and ``best of three strategy''~\cite{10.1007/978-3-031-48550-3_17} can help in improving the consistency of the LLM output, which was a major limitation mentioned by the experts related to requirements analysis. The reasoning guidelines such as chain-of-thought and tree-of-thought can help in improving the reasoning capabilities of the LLMs, which is another limitation emphasised by the experts within this phase. 

For users performing requirements specification using LLMs can benefit from using ``Persona'' and ``Disambiguation'' templates. The primary reason is eliminate and reduce ambiguities present within requirements specifications as per the experts' opinions.

For requirements validation using LLMs, ``persona'' guidelines can be helpful for validating requirements against the needs of specific user types. The ``template'' guidelines can help in having some predefined approaches which saves time by reducing the need for experimenting with finding optimal prompts for particular use cases.

Finally, for users performing requirements management tasks using LLMs, ``keyword'' and ``reasoning'' guidelines can be most helpful. Keyword guidelines can help put the task being performed into appropriate contexts by establishing the right assumptions. The need for traceability and and reasoning can be addressed by following ``reasoning'' guidelines while prompting LLMs.

\section{Discussion} \label{sec:discussion}

One of the main objectives of this study was to explore how existing prompt engineering guidelines can help users in leveraging LLMs in RE activities. Our findings indicate that while various guidelines exist, they originate from a diverse range of models, including text-to-text models such as GPT-3.5 and BERT, as well as text-to-image and multi-modal models like DALL-E and the combination of CLIP and VQGAN. Although some guidelines exhibit overlap, our analysis reveals notable differences, suggesting that the role of prompt engineering will continue to grow in importance. Notably, the majority of guidelines identified through our literature review were mentioned only once across studies. This shows that, while there is much interest in improving the performance of the LLM through coming up various prompt based approaches, this field of study is still in its early stages and needs to mature before this becomes an established discipline of study. The guidelines and techniques need to be validated across broader range of use cases to reach a generalisable conclusion on which techniques work reliably enough for which tasks and use cases and what are the constraints and limitations of them.

We also examined how these guidelines apply to different RE activities by mapping them to guideline themes. Our interviews confirmed that while multiple interpretations of these mappings exist, there was general agreement on the value of the guideline themes. The template theme guidelines seems to the most relevant ones, as they are mapped to three of the five RE activities. The keyword, persona and reasoning theme guidelines were mapped to two activities each while the context, analysis and disambiguation themes were only mapped to one activity. However, based on the experts' insights, while the context and disambiguation guidelines are applicable to fewer use cases, they bring more value to the use cases they are applicable to compared to the rest of the guideline themes. This indicates that the prompt engineering guidelines can serve two major purposes: broadly apply to various RE activities and bring value to the tasks they are used within.

Another interesting observation is that these seven themes of prompt engineering guidelines make up ~72\% of the all the prompt engineering guidelines we found. Some guidelines, such as those focused on few-shot prompting and wording which make up ~28\% of all the prompt engineering guidelines we found in our literature, were deemed irrelevant to RE and were not part of the mapping. This indicates that, while existing prompt engineering guidelines are fairly applicable to RE activities, there is also a risk that a good chunk of them are not. This points to the need for either developing new prompt engineering guidelines, specific to RE or fine-tuning existing guidelines to make them more applicable and effective for RE. Overall, our findings show both the potential and limitations of prompt engineering guidelines for using LLMs in RE. Providing context is key to improving LLM performance, but more research is needed to refine these guidelines and explore their use in more advanced RE activities.

\subsection{Threats to Validity}

While we followed established methods and techniques for the collection and analysis of the data to ensure methodological rigour and soundness, it is essential to acknowledge and address potential threats to the validity of the findings.

\textbf{Internal Validity:} Despite selecting a review period from 2018 to the present, there is a risk of omitting relevant studies. However, as large language models (LLMs) largely emerged after 2017, and prompt-related guidelines first appeared in 2021, this risk is likely minimal. Limited available literature and inconsistencies in terminology pose risks. Selection bias may arise from reviewing a restricted set of studies. Open discussions among researchers aimed to minimise misinterpretations.

\textbf{External Validity:} The study’s findings may have limited generalisability due to its specific focus and the rapidly evolving field. Including unpublished preprints from arXiv introduces potential bias but helps capture cutting-edge research. Interviewed experts may have personal biases influencing their responses, potentially introducing confirmation bias.

\textbf{Construct Validity:} The study’s inclusion and exclusion criteria may affect the construct validity. Also, the interviewed RE experts' varying understanding of the terminology within the guidelines is another threat.

\section{Conclusion} \label{sec:conclusion}

The primary motivation to conduct and present this study was to address the lack of domain specific prompt engineering guidelines for using LLMs, especially within RE. To that extent, our study provided an overview of existing prompt engineering guidelines. We examined the benefits and limitations of these guidelines through interviews with three RE experts. Our interviews highlighted the usefulness of the identified guidelines, especially for the context, template, keywords, analysis, reasoning, persona, and disambiguation guidelines for various RE activities. Based on the insights gathered from these interviews, we proposed a mapping for prompt engineering guidelines for different RE activities.

We believe that as LLMs become more reliable in RE activities, they could help reduce errors, detect issues earlier in the process, and prevent costly corrections later in development. This, in turn, could lower project failure rates and improve overall efficiency in RE workflows. Our findings contribute to the improvement of RE practices by supporting the development of tailored prompt engineering guidelines. Future work should explore how general prompt guidelines can be adapted for specific domains like RE and whether additional fine-tuning of these guidelines is required to make them more effective.

\bibliographystyle{plainnat}
\bibliography{reference}

\end{document}